\documentclass[twocolumn,usenatbib,useAMS,fleqn,twocolappendix]{aastex63} 
\usepackage{natbib}
\usepackage{amsmath}
\usepackage{amssymb}
\usepackage{graphicx}
\usepackage{subfigure}
\usepackage{color}
\usepackage{epsfig}
\usepackage{cancel}
\definecolor{internationalkleinblue}{rgb}{0.0, 0.18, 0.65}
\definecolor{britishracinggreen}{rgb}{0.0, 0.26, 0.15}
\hypersetup{colorlinks=true,
citecolor=internationalkleinblue, linkcolor=internationalkleinblue,
urlcolor=internationalkleinblue}

\received{\today}
\submitjournal{ApJL}

\newcommand{\vrel}{v_{\rm rel}}

\newcommand{\vturb}{v_{\rm turb}}
\newcommand{\vturbL}{v_{\rm turb, L}}
\newcommand{\Lturb}{L_{\rm turb}}

\newcommand{\vin}{v_{\rm in}}

\newcommand{\cs}{c_{\rm s}}
\newcommand{\cshot}{c_{\rm s,hot}}

\newcommand{\rhoh}{\rho_{\rm hot}}
\newcommand{\rhoc}{\rho_{\rm cold}}

\newcommand{\Thot}{T_{\rm hot}}
\newcommand{\Tcold}{T_{\rm cold}}
\newcommand{\Tmix}{T_{\rm mix}}
\newcommand{\tc}{t_{\rm cool}}
\newcommand{\tmix}{t_{\rm mix}}
\newcommand{\tsh}{t_{\rm sh}}

\newcommand{\tstc}{t_{\rm sh}/t_{\rm cool}}

\newcommand{\mach}{\mathcal{M}}
\newcommand{\macht}{\mathcal{M}_{\rm turb}}

\newcommand{\eg}{e.g.,}

\newcommand{\ft}{f_{\rm turb}}

\newcommand{\edotcool}{\dot{E}_{\rm cool}}
\newcommand{\mdot}{\dot{M}}
\newcommand{\pdot}{\dot{p}}

\shorttitle{Fractal Mixing Layers}
\shortauthors{Fielding et al.}
\begin{document}

\title{Multiphase Gas and the Fractal Nature of Radiative Turbulent Mixing Layers}

\correspondingauthor{Drummond B. Fielding}
\email{drummondfielding@gmail.com}

\author[0000-0003-3806-8548]{Drummond B. Fielding}
\affiliation{Center for Computational Astrophysics, Flatiron Institute, 162 5th Ave, New York, NY 10010, USA}

\author[0000-0002-0509-9113]{Eve C. Ostriker}
\affiliation{Department of Astrophysical Sciences, Princeton University, Princeton, NJ 08544, USA}

\author[0000-0003-2630-9228]{Greg L. Bryan}
\affiliation{Department of Astronomy, Columbia University, 550 W 120th Street, New York, NY 10027, USA}
\affiliation{Center for Computational Astrophysics, Flatiron Institute, 162 5th Ave, New York, NY 10010, USA}

\author[0000-0001-5048-9973]{Adam S. Jermyn}
\affiliation{Center for Computational Astrophysics, Flatiron Institute, 162 5th Ave, New York, NY 10010, USA}

\begin{abstract}
A common situation in galactic and intergalactic gas involves cold dense gas in motion relative to hot diffuse gas. 
Kelvin-Helmholtz instability creates a turbulent mixing layer and populates the intermediate-temperature phase, which often cools rapidly.
The energy lost to cooling is balanced by the advection of hot high enthalpy gas into the mixing layer, resulting in growth and acceleration of the cold phase. 
This process may play a major role in determining the interstellar medium and circumgalactic medium phase structure, and accelerating cold gas in galactic winds and cosmic filaments. 
Cooling in these mixing layers occurs in a thin corrugated sheet, which we argue has an area with fractal dimension $D=5/2$ and a thickness that adjusts to match the hot phase mixing time to the cooling time. 
These cooling sheet properties form the basis of a new model for how the cooling rate and hot gas inflow velocity depend on the size $L$, cooling time $\tc$, relative velocity $\vrel$, and density contrast $\rhoc/\rhoh$ of the system. 
Entrainment is expected to be enhanced in environments with short $\tc$, large $\vrel$, and large $\rhoc/\rhoh$.
Using a large suite of three dimensional hydrodynamic simulations, we demonstrate that this fractal cooling layer model accurately captures the energetics and evolution of turbulent interfaces and can therefore be used as a foundation for understanding multiphase mixing with strong radiative cooling.

\end{abstract}
\section{Introduction} \label{sec:intro}

Prevalent on nearly all scales within and around galaxies is the presence of colder gas moving relative to hotter ambient material. Often the cold and hot phases are in pressure and thermal equilibrium (or negligibly cooling) and 
mixing at the interfaces driven by Kelvin-Helmholtz instabilities (KHI) populates the thermally unstable intermediate temperature phase. These radiative mixing layers are essential in setting the phase structure in the interstellar medium (ISM) \citep{Audit+10, Kim+13}, circumgalactic medium (CGM) \citep{Fielding+17,Ji+19}, and intracluster medium (ICM) \citep{Gaspari+12, Banerjee+14,YuanLi+19}, and regulate the evolution of supernova remnants and superbubbles \citep{Kim+17,Fielding+18, ElBadry+19}, cosmic filaments \citep{Mandelker+19b}, galactic winds \citep{Gronke+20}, protoplanetary disk dynamics, and protostellar (and potentially active galactic nuclei) jets \citep{Stone+97}. The underlying physics is analogous to the opposite problem of burning/energy release in turbulent media, which takes place in stellar interiors, supernovae, and rocket engines \citep[\eg][]{NiemeyerKerstein+97}. Moreover, there are close parallels to physical processes in planetary clouds where energy is exchanged via phase change instead of radiation \citep{Pauluis+11}. 

Understanding radiative mixing layers is crucial to theories of galaxy formation and evolution  because these layers can dominate the energetics and regulate the amount of cold gas available for star formation. They are, therefore, also essential for connecting to observations of gas in and around galaxies, which are most sensitive to cooler gas phases rather than hot dilute gas. In particular, recent observations of galactic winds \citep[\eg][]{Heckman+15, Chisholm+17, McQuinn+19} and the CGM \citep[\eg][]{Prochaska+17, Rubin+18, Rudie+19, Zahedy+19} have challenged simulations and theories with constraints on the kinematics, sizes, metallicities, and broad range of temperatures in these systems. More generally, the prevalence of multiphase gas in many observed systems begs the question: \emph{how are energy, mass, and momentum transferred between the hot and cold phase in different environments?}

This question has been studied in various guises. In the context of ISM bubbles and clouds, the competition of conduction, cooling, and/or turbulent mixing is a long standing question \citep[\eg][]{Cowie1977,McKeeCowie77, ElBadry+19}. Many simulations have focused on cloud-crushing, acceleration, and destruction by a hot, high-velocity flow \citep[\eg][]{Klein1994, Scannapieco+15, Schneider+17}, and there is evidence that thermal instability and mixing aids in the development and persistence of the CGM and ICM cold phase \citep[\eg][]{McCourt+12,Voit18,Prasad+18}.

Radiative mixing layers are an inherently small scale process, which makes accurately capturing their impact on global scales challenging. Recent attempts to better resolve the CGM cold phase in cosmological contexts have demonstrated the impact of inadequate resolution on observational predictions and simulated galaxy properties \citep{vandeVoort+19, Hummels+19, Peeples+19}. Fully resolving from the halo scale (100s kpc) down to the cold gas scale (0.1-10 pc; \eg\, \citealt{McCourt+18, Gronke+20}) may be necessary to resolve apparent discrepancies, such as the vastly higher galactic wind mass outflow rates needed by cosmological simulations \citep[\eg][]{Nelson+19} compared to what is predicted by simulations of the star-forming ISM \citep[\eg][]{Kim+18}. 
These resolution requirements are daunting, and they motivate our search for an effective theory of radiative mixing layers that could be used to model the smallest scales.

\cite{BegelmanFabian90} presented a model for radiative mixing layers in which cooling is balanced by the advection of high enthalpy hot gas with assumed energy flux $\sim P \vturb$. \cite{ElBadry+19} analyzed quasi-steady diffusive mixing/cooling interfaces and showed that the energy flux is $\sim P (\kappa_{\rm diff}/\tc)^{1/2}$ where $\kappa_{\rm diff}$ is the effective diffusivity and $\tc$ the cooling time of intermediate-temperature gas. Recent numerical simulation studies of strongly cooling turbulent mixing layers have found that the cold phase grows when the cooling time of the mixed gas is shorter than the mixing time, and does so at a rate $\propto \tc^{-1/4}$ \citep{Gronke+18, Gronke+20, Mandelker+19b}. While the above work addressed important aspects of turbulent mixing/cooling layers, a complete physical model has not previously been formulated.

In this \emph{Letter}, we employ analytic arguments and numerical simulations to investigate turbulent mixing layers with radiative cooling, considering a wide range of parameters. We begin in \autoref{sec:model} by developing a new model that explains the total cooling, growth rate, and acceleration of the cold phase by considering the enthalpy flux through the fractal surface that delineates the strongly cooling layer. In \autoref{sec:experiment} and \autoref{sec:results} we describe our numerical experiment design and results, respectively, which provides strong support for our theory. 

In a forthcoming companion paper, henceforth referred to as Paper II (Fielding et al., in prep), we delve deeper into the details of the theoretical basis and experimental evidence for the results presented here.

Movies of our simulations can be found at \url{https://dfielding14.github.io/movies/}.

\section{Fractal Cooling Layer Model} \label{sec:model}

Consider the most general form of a radiative turbulent mixing layer in which cold and hot gas in pressure and thermal equilibrium move relative to each other. The KHI quickly develops turbulence that promotes mixing and populates the rapidly cooling intermediate temperature phase. Some of the astronomical applications we have in mind are a dense clump being enveloped by a supernova remnant, a cold cloud being ablated by a hot wind, a cold blob moving relative to a hot CGM, or a cosmic filament flowing into a gaseous halo, but we keep our formulation general to allow our model to be applied to a broad range of scenarios. 

The evolution of the system is controlled by three dimensionless numbers, which are
\begin{subequations}
\begin{align}
\xi &= \tsh/\tc = L / (\vrel \tc)\\
\chi &= \rhoc/\rhoh \\
\mach &= \vrel/\cshot,
\end{align}
\end{subequations}
where $\vrel$ is the relative velocity of the hot and cold phases, $L$ is the characteristic streamwise length of the mixing layer, $\tsh = L/\vrel$ is the shear time, $\tc$ is the minimum cooling time, which generally occurs at intermediate temperatures, $\rhoc$ and $\rhoh$ are the cold and hot phase densities, and $\cshot$ is the hot phase sound speed.   

In quasi-steady state in the frame of the interface, radiative cooling losses are balanced by the advection of hot high enthalpy gas. Hot gas flows into the cooling layer at a speed $\vin$ carrying mass and momentum. The inflow velocity $\vin$, therefore, encapsulates the total cooling rate, the mass transfer rate from hot to cold, and the transport rate of momentum (producing cold phase acceleration): 
\begin{subequations}
\begin{align}
  \edotcool &\approx (E_{\rm th} + P) L^2 \vin \label{eq:edot} \\
  \mdot &\approx \rhoh L^2 \vin \label{eq:mdot} \\
  \pdot &\approx \rhoh \vrel L^2 \vin  \label{eq:pdot}.  
\end{align}
\end{subequations}  
The balance between the advected enthalpy flux and the radiative losses integrated over the volume gives an expression for $\vin$:
\begin{align}
\int {\bf v} \cdot \nabla \left( \tfrac{\gamma}{\gamma - 1} P\right) dV &= \int \dot{\mathcal{E}}_{\rm cool} dV \nonumber \\
\Rightarrow \quad \frac{\gamma}{\gamma - 1} P \vin L^2 &= \frac{\gamma}{\gamma - 1} \frac{P}{\tc} w A_w \nonumber \\
\Rightarrow \quad \frac{\vin}{\vrel} &=\frac{\tsh}{\tc} \left(\frac{w}{L}\right) \left(\frac{A_w}{L^2}\right),
\label{eq:vin}
\end{align}
where $w$ and $A_w$ are the thickness and area of the thin sheet where cooling takes place. It is essential to realize that $A_w \gg L^2$ because this sheet is highly corrugated. Here we have assumed that the cooling is isobaric, and that the cooling is dominated by the gas that cools with cooling time $\tc$, which is supported by our simulations. 

The characteristic cooling layer thickness $w$ is set by the length scale on which hot gas is mixed in at the same rate that it cools. The hot mixing rate can be estimated using the fact that the turbulent velocity of these flows is subsonic, so the turbulent energy densities of the hot and cold phase are nearly equal\footnote{In Paper II we will demonstrate that the amount of work done on the turbulent field by cooling is small.}. Hence, $\rhoh v_{\rm turb,hot}^2 = \rhoc v_{\rm turb, cold}^2$ or $v_{\rm turb,hot} = \chi^{1/2} v_{\rm turb, cold}$. 
For concise notation we define $\vturb \equiv v_{\rm turb, cold}$. Putting this together we can estimate the cooling layer thickness $w$ using
\begin{align}
\tmix(w) &= \frac{w}{v_{\rm turb,hot}(w)}= \frac{w}{\chi^{1/2} \,  \vturbL \left(\frac{w}{L}\right)^{1/3}} = \tc \nonumber \\ 
\Rightarrow \quad \frac{w}{L} &= \left(\frac{\tc}{\tsh}\right)^{3/2} \left(\frac{\vturbL}{\vrel} \right)^{3/2} \chi^{3/4},
\label{eq:w}
\end{align}
where $\vturbL$ is the turbulent velocity on the scale $L$, and the second equality relies on the subsonic Kolmogorov turbulent velocity structure function, $\vturb(\ell)= \vturbL (\ell/L)^{1/3}$, to estimate the characteristic turbulent velocity on a given scale.

The magnitude of $\vturbL$ in the fully non-linear state depends only on $\vrel$ with a weak time dependence. In Paper II we shall present theoretical and empirical evidence for this fact, but this should be intuitively understandable because the only source of free energy to drive the turbulence is the shear velocity (the free energy in the thermal energy gradient is inaccessible because the flow is subsonic). 
We shall define $\ft \equiv {\vturb}/{\vrel}$, which from our numerical experiments typically takes on a value ${\sim} 0.1{-}0.2$.
This agrees with previous, albeit non-radiative, shear flow studies \citep{Mandelker+19a}.

The cooling layer area $A_w$ can be estimated by utilizing the fractal nature of the surface. Specifically, the fractal dimension provides a measure of the scale dependent surface area. The area of a non-fractal surface (e.g., a sphere, or cube) scales with the square of the linear size of the object $L^2$ and is independent of the measurement scale. By contrast, the area of a fractal surface (e.g., a coastline, cauliflower, or ball of crumpled paper) scales with the size of the object to a larger, usually non-integer, power, which depends on the measurement scale. We let $D$ be the fractal dimension so that $d=D-2$ is the excess dimensionality over a non-fractal scaling. In this convention $A_\lambda/L^2 = (L/\lambda)^d$ for measurement scale $\lambda$ \citep{Sreenivasan+89}.

We can predict the fractal dimension by analogy to well-known fractals. The cooling surface can be approximated by large mode sinusoidal perturbations with successively smaller modes on top. This is reminiscent of the Koch curve/surface that is constructed by iteratively deforming a flat line/surface up on one side and down on the other with two squares/cubes. The Koch curve and surface have $d=1/2$. The $d=1/2$ may also be understood by noting that the turbulent velocity field tends to perturb the cooling surface up or down, and nearby regions will be correlated. This is similar to a regular Brownian surface on which the average height difference between two points scales with the square of the distance, which also has a fractal dimension corresponding to $d=1/2$. Moreover, it has been shown empirically and predicted theoretically that isocontours in compressive turbulence have fractal dimensions corresponding to $d=1/2$ \citep{Mandelbrot75,Federrath+09}. Although the turbulence in radiative mixing layers is subsonic, the compressive nature of cooling will change the flow dynamics. We, therefore, adopt $d=1/2$, or
\begin{align}
\frac{A_\lambda}{L^2} = \left(\frac{\lambda}{L}\right)^{-1/2}.
\label{eq:area}
\end{align}
This relation is expected to hold for all scales $\lambda$ that are greater than the dissipative scale and smaller than $L$. In the limit of strong cooling and weak dissipation this area relation applies to the cooling layer area $A_w$.

We now return to \autoref{eq:vin} and plug in our predictions for the thickness $w$ and area $A_w$ of the cooling layer from \autoref{eq:w} and \autoref{eq:area} respectively to obtain the expression
\begin{subequations}
\begin{align}
\frac{\vin}{\vrel} &= \chi^{3/8} \xi^{1/4} \ft^{3/4} \label{eq:final}
 \\ 
\frac{\vin}{\cshot} &= \chi^{3/8} \xi^{1/4} \mach \ft^{3/4}.
\end{align}
\end{subequations}
This simple power-law expression for the inflow velocity, and so also $\edotcool$, $\mdot$, and $\pdot$, encapsulates the essential behavior of radiative mixing layers in terms of the three characteristic dimensionless parameters that describe the bulk properties. 

Although this model has been formulated specifically for systems where shear flows lead to turbulence and then to mixing and cooling, it should apply equally well for systems in which turbulence has an alternative driving mechanism. Hence in general we expect the hot gas inflow velocity to obey 
\begin{align}
\vin = C \left(\frac{\rhoc}{\rhoh}\right)^{3/8} \left(\frac{\Lturb}{\tc}\right)^{1/4} \vturbL^{3/4}, \label{eq:general}
\end{align}
where $\vturbL$ is the turbulent velocity on the outer scale $\Lturb$ of the turbulence, and $C$ is a constant dependent on the exact geometry of the problem and what is driving the turbulence (\eg \,Rayleigh-Taylor instability or cloud-crushing).

\begin{figure*}
\includegraphics[width = \textwidth]{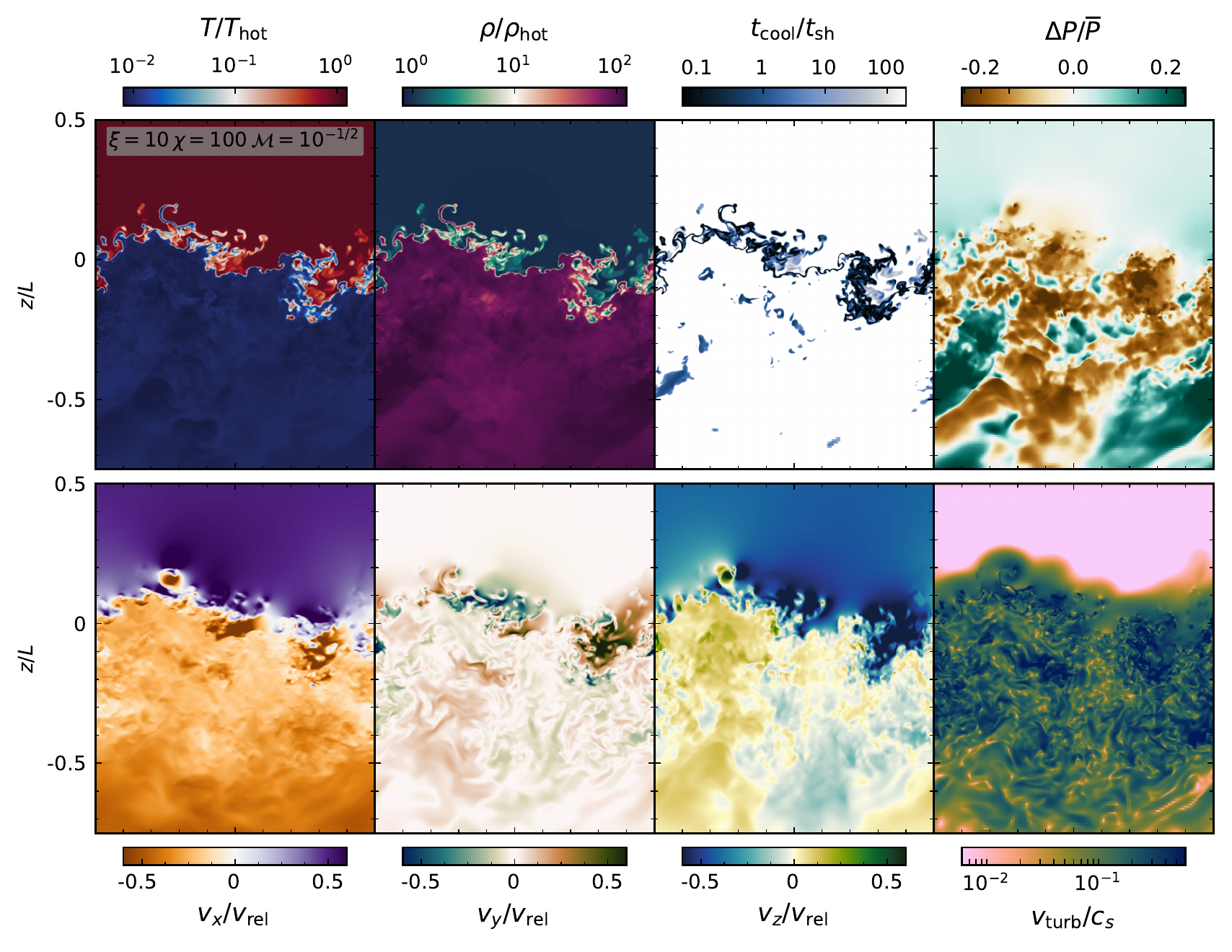}
\caption{ From left to right and top to bottom, slices of temperature, density, cooling time, pressure deviation, $v_x$, $v_y$, $v_z$, and turbulent Mach number at $t/\tsh=30$ for a $\xi=10, \, \chi=100, \, \mach=10^{-1/2}$ simulation. The background shear flow is in the $\hat{x}$ (horizontal) direction, with the hot gas  moving to the right relative to the cold. The turbulence, traced by $v_y$, has induced mixing and broadened the shear velocity $v_x$, but the rapid cooling, localized entirely to a thin layer, maintains a sharp gradient between the cold and hot phases. The cooling kindled by the mixing also leads to a flow of the hot gas into the cooling layer, $v_z < 0$. Although the cooling is rapid, there is no signature of the cooling imprinted in the pressure field; instead the pressure fluctuations correlate with turbulent fluctuations. An animated version of this figure is available \textsf{\href{https://vimeo.com/397632983}{here}}. \label{fig:map}} 
\end{figure*}

\begin{figure*}
\includegraphics[width = 0.5\textwidth]{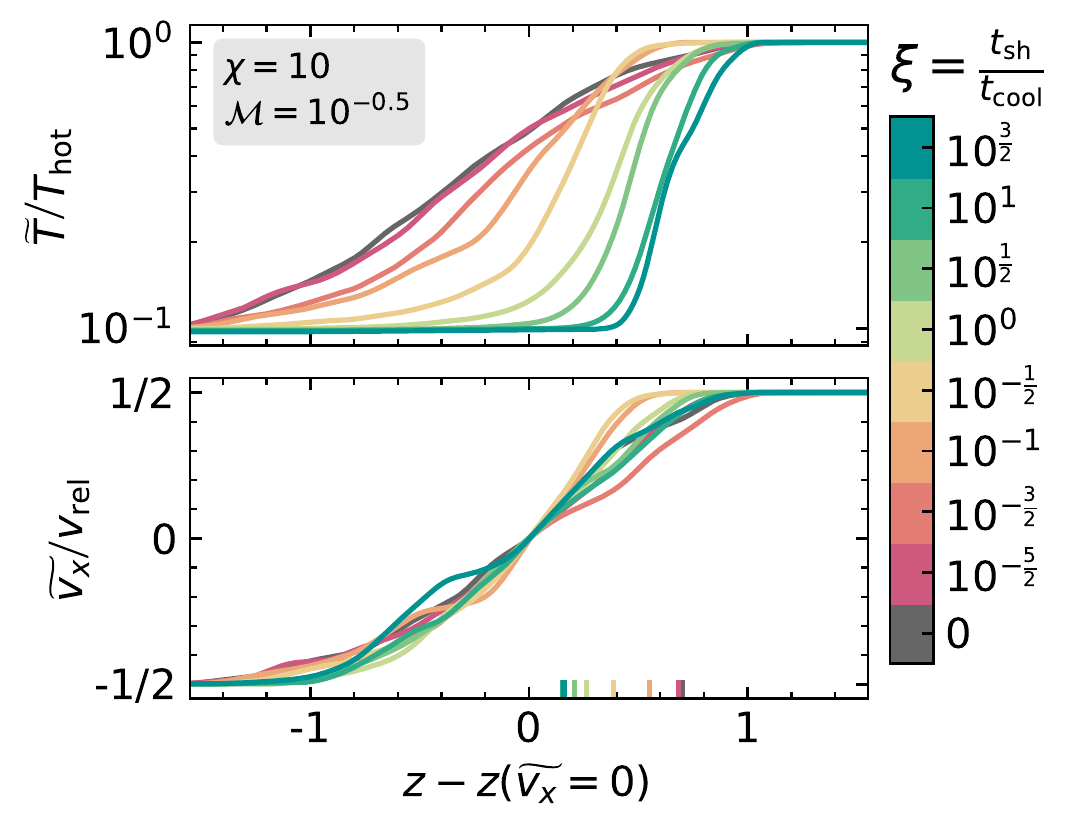}
\includegraphics[width = 0.5\textwidth]{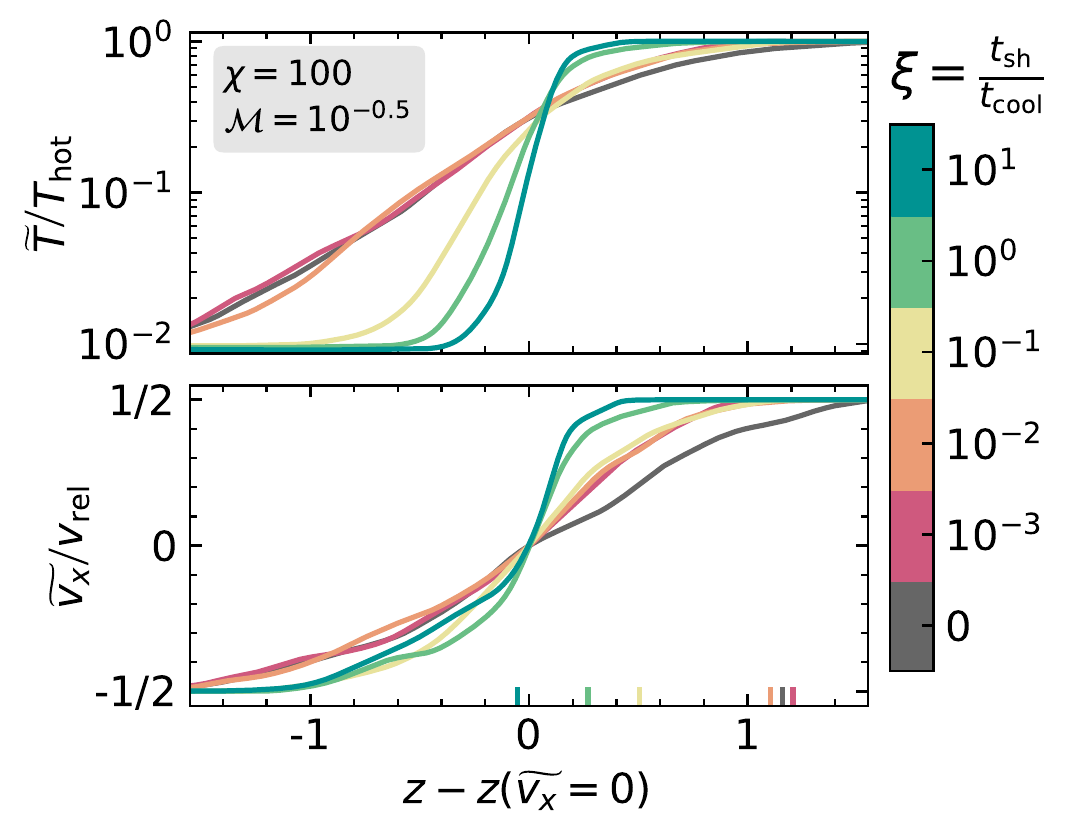} 
\caption{ Mass-weighted horizontally averaged temperature $\widetilde{T}$ (top) and shear velocity $\widetilde{v_x}$ (bottom) profiles at $t/\tsh = 30$ for simulations spanning a wide range of $\xi$ values with $\chi=10$ (left) and $\chi=100$ (right). The profiles have been shifted so $\widetilde{v_x} = 0$ at the same point. The colored ticks indicate $z(\widetilde{v_x} = 0)$. Adiabatic and slowly cooling simulations ($\xi \ll 1$) have broad $\widetilde{T}$ and $\widetilde{v_x}$ profiles. As cooling increases the $\widetilde{T}$ profile gets steeper, but $\widetilde{v_x}$ stays nearly the same, highlighting the difference between the thermal and momentum mixing layers. \label{fig:profiles}}
\end{figure*}

\begin{figure*}
\includegraphics[width = 0.5\textwidth]{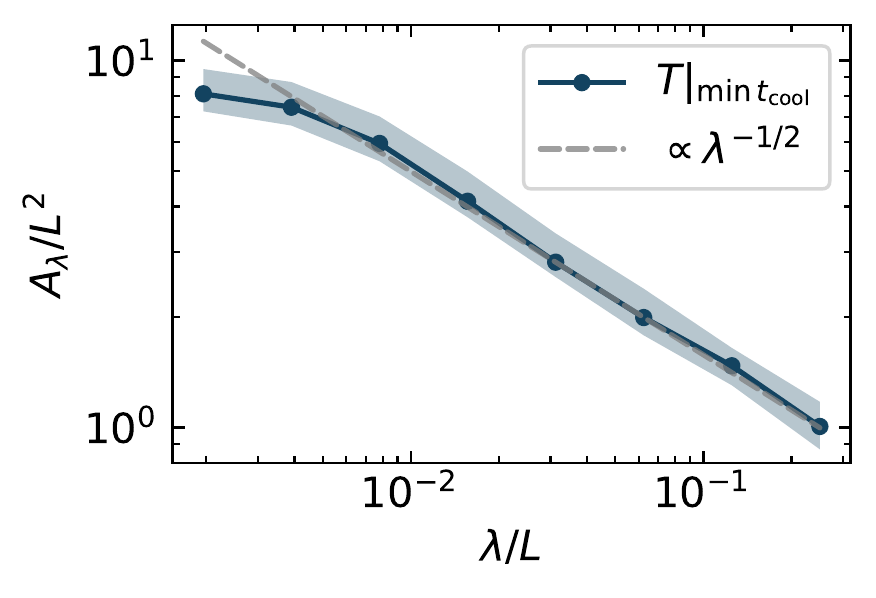}
\includegraphics[width = 0.5\textwidth]{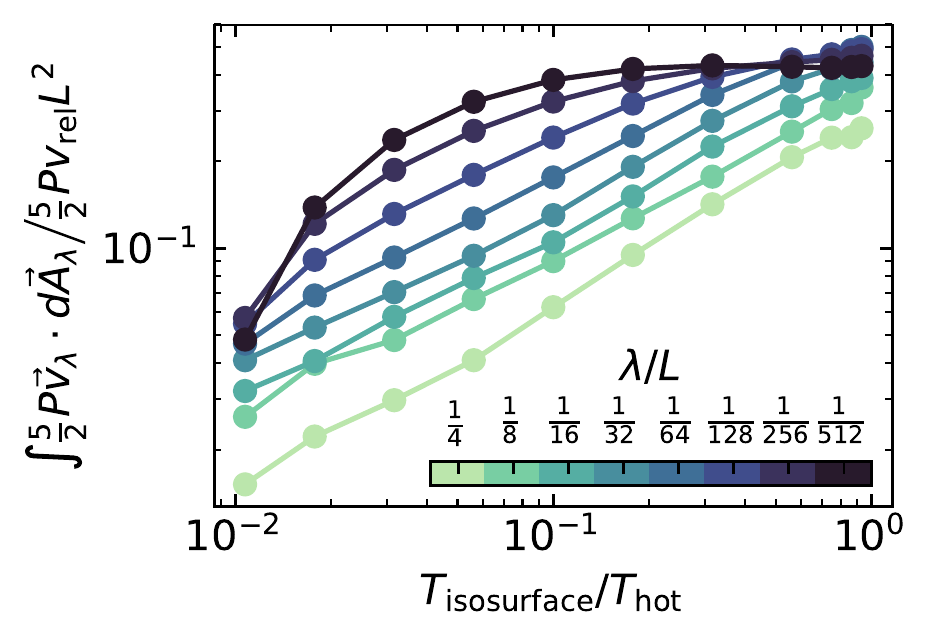}
\includegraphics[width = \textwidth]{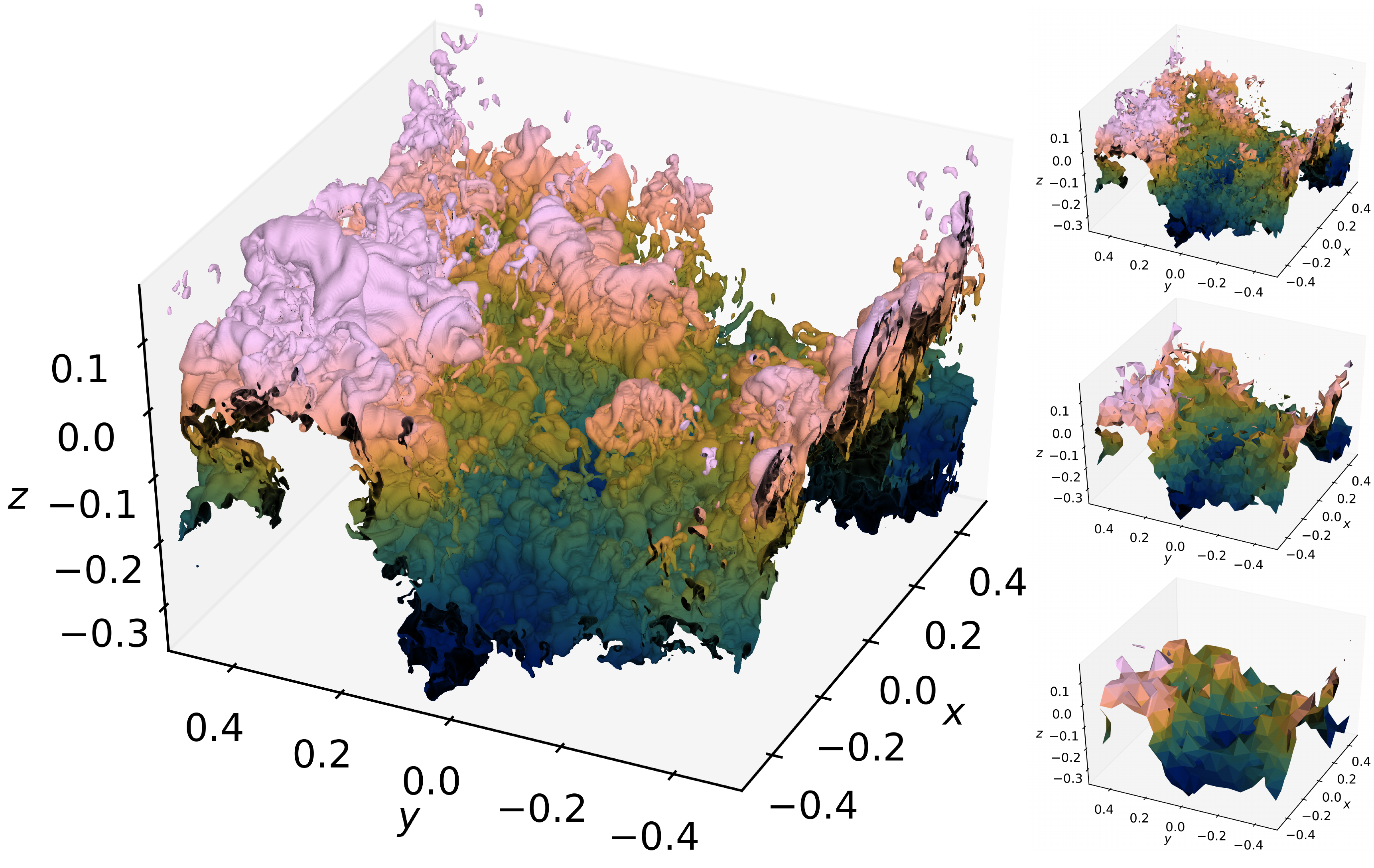}
\caption{The fractal nature of the cooling surface in the same exemplary simulation as in \autoref{fig:map} which has $\xi=10, \, \chi=100,\,\mach=10^{-1/2}$, and $\Delta x = L/512$. The {\bf lower left} panel shows the $T$ isosurface where $\tc$ is minimized. The color denotes the height. The apparent variations on all scales is indicative of the fractal nature of the surface. The area of the isosurface decreases when the temperature field is blurred on scale (i.e., downsampled by a factor of) $\lambda$. This is shown pictorially in the small {\bf lower right} panels which show, from top to bottom, the surface when blurred on scale $\lambda = 8,\, 16$ and $32\, \Delta x$ = $L/64,\, L/32,$ and $L/16$, respectively. The {\bf top left} panel shows quantitatively how the area changes with the blurring scale $\lambda$. The shaded region shows the 1 $\sigma$ temporal variations. The logarithmic slope of the $A_\lambda$ relation is very well fit by $A_\lambda \propto \lambda^{-1/2}$, which corresponds to a fractal dimension of $D=2.5$, $d=1/2$. The {\bf top right} panel shows the thermal energy flux through isosurfaces defined at a range of temperatures when blurred to varying degrees. The curves for the least blurred isosurfaces (darkest) demonstrate that the thermal energy flux is constant until cooling kicks in at $T\lesssim \Tmix$. An animated version of this figure is available \textsf{\href{https://vimeo.com/398055547}{here}}.	\label{fig:Fractal_Dimension}}
\end{figure*}

\begin{figure}
\hfill\includegraphics[width = 0.95\columnwidth]{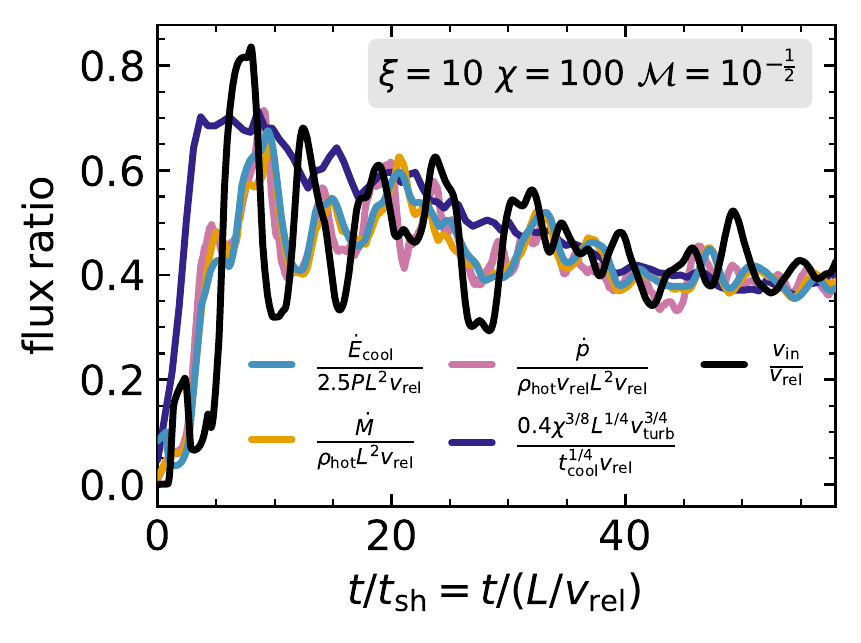}\\
\includegraphics[width = \columnwidth]{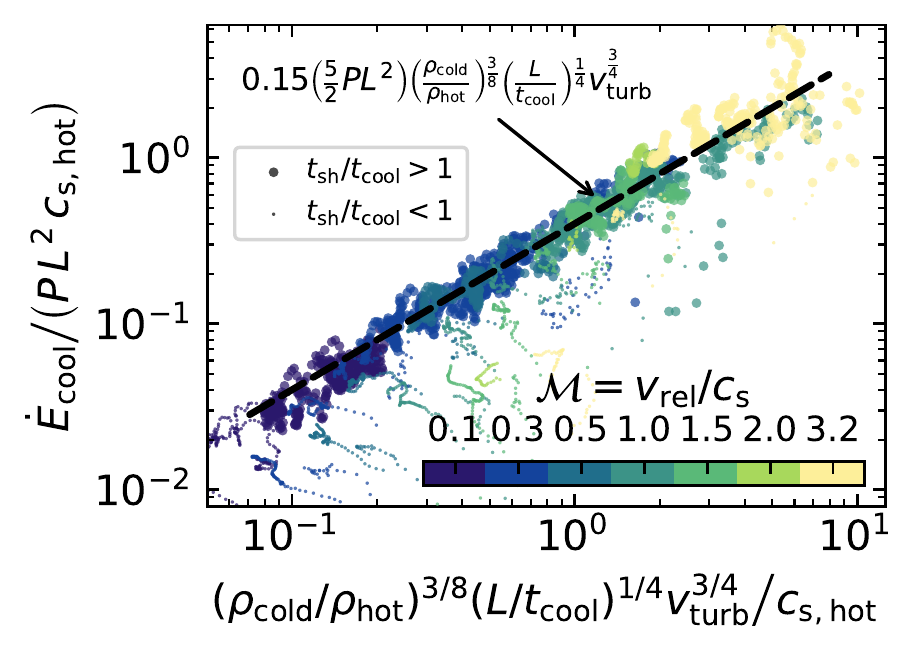}\\
\includegraphics[width = \columnwidth]{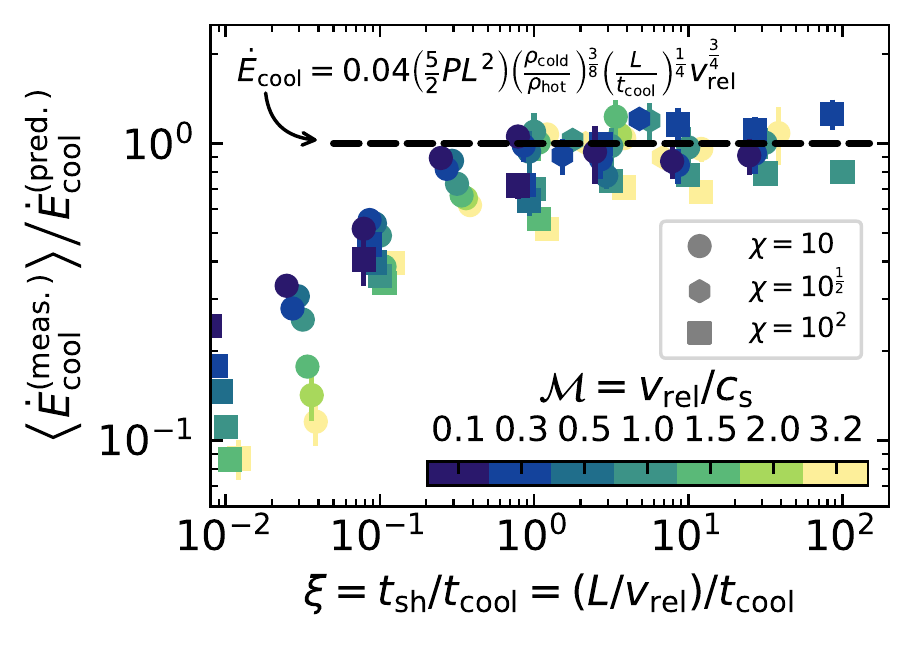}
\caption{({\bf Top}) The inflow velocity (black), cooling rate (blue), mass flux (gold), and momentum flux (pink) with normalizing factors for each quantity as shown. These match over time in the fiducial simulation and coincide closely with the predicted value (purple, \autoref{eq:final}).
({\bf Middle}) The instantaneous cooling rate at all times for all simulations versus the predicted scaling, demonstrating that when cooling is rapid $(\tstc > 1$; large points) the fractal cooling layer model holds. The slowly cooling systems $(\tstc < 1$; small points) have yet to reach, but are approaching, the equilibrium relation.
({\bf Bottom}) Average cooling rate (and $2\sigma$ variation) normalized by the predicted enthalpy flux  over $20-40 \, \tsh$ for all simulations (\autoref{eq:final}). We adopt a coefficient of $0.04$
that includes $\ft\approx0.15$ and the order unity constants in the $w$ and $A_w$ definitions. \label{fig:Edot}} 
\end{figure}

\begin{figure*}

\includegraphics[width = 0.495\textwidth]{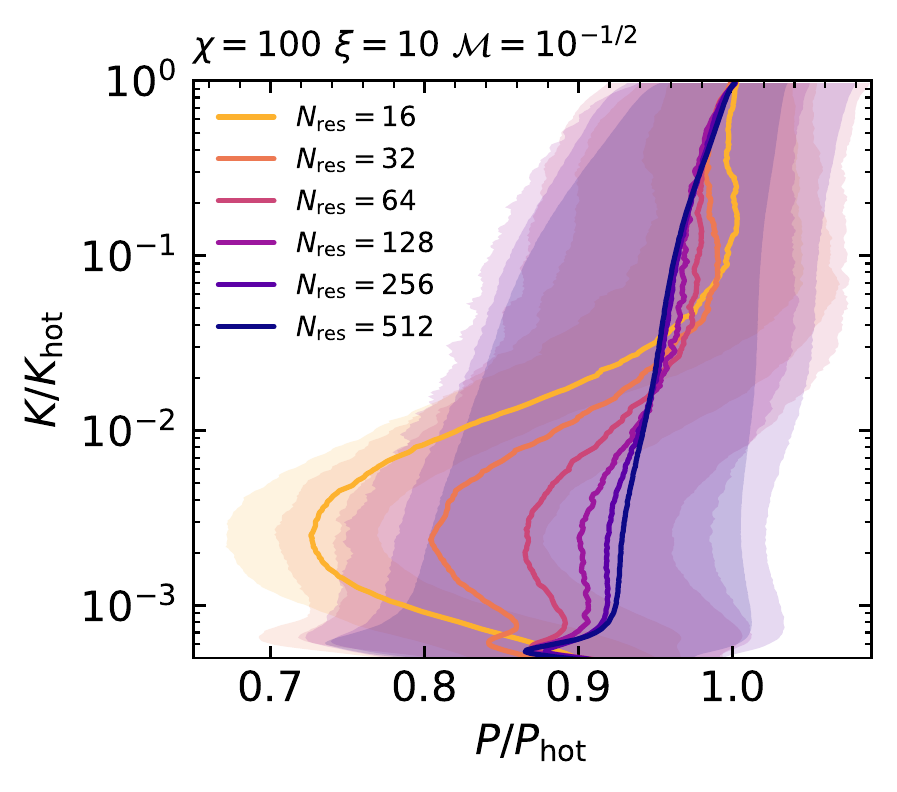} 
\includegraphics[width = 0.495\textwidth]{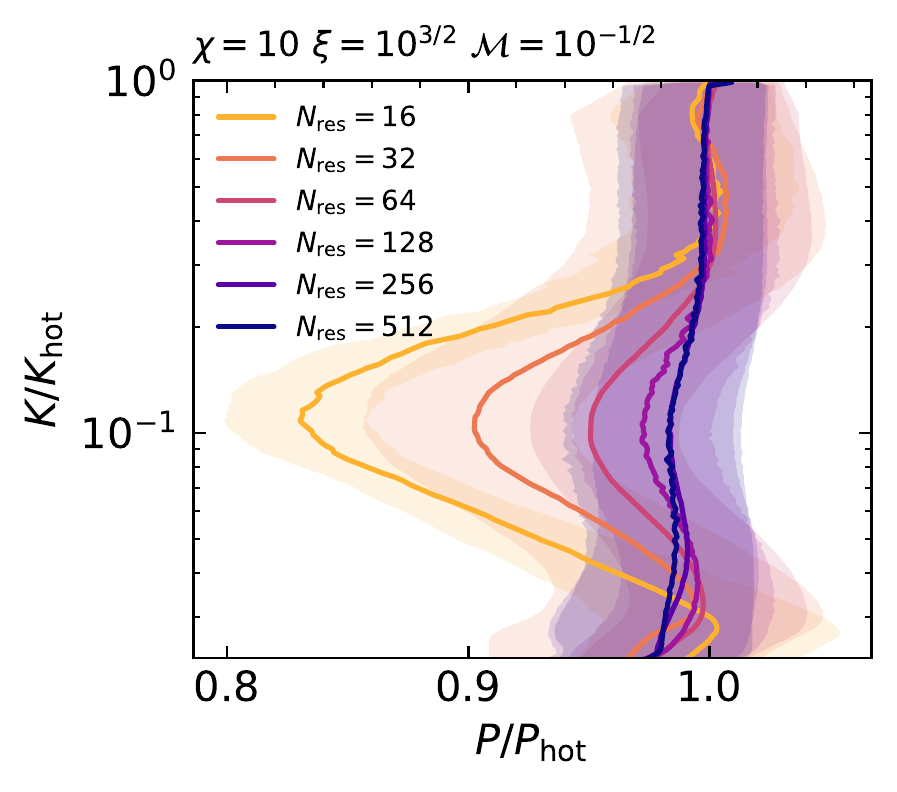}\\
\includegraphics[width = 0.5\textwidth]{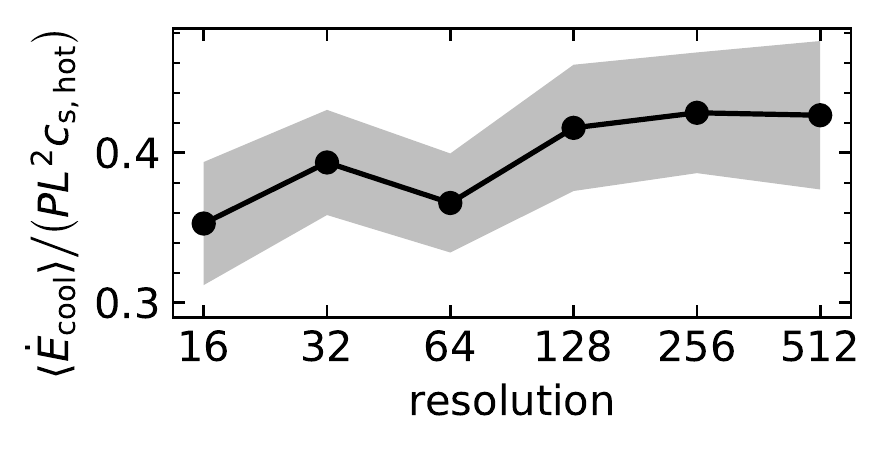}
\includegraphics[width = 0.5\textwidth]{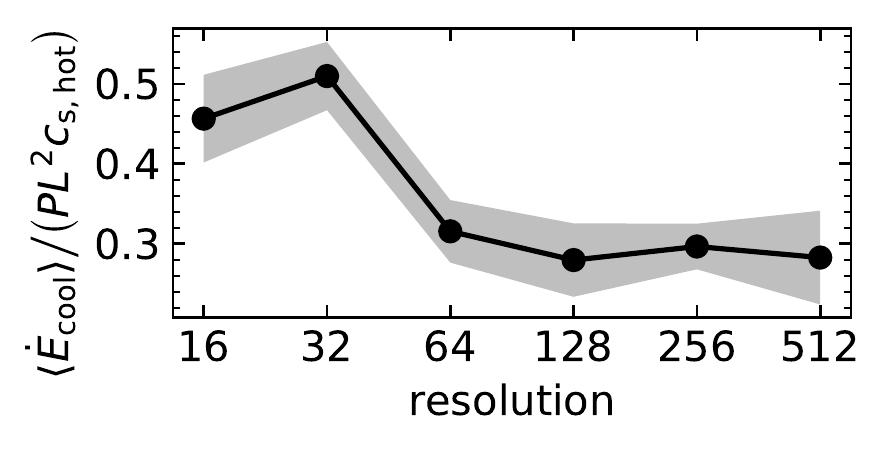}
\caption{({\bf Top}) The median and 1 $\sigma$ pressure-entropy mass distribution for two choices of dimensionless parameters at resolutions ranging from 16 to 512 elements per $L$. The low resolution simulations exhibit pressure decrements of up to 30 percent at low/intermediate entropies where the cooling rate peaks, while the converged higher resolution simulations cool isobarically. 
({\bf Bottom}) The average and 1 $\sigma$ variation of the cooling rate as a function of resolution demonstrates the cooling rate convergence at high resolution ($\Delta x \lesssim L/128$).  Although the lower resolution simulations are accurate to within a factor of $\lesssim 2$
of the converged value at $N_{\rm res} \gtrsim 128$,
the offset can go either way.
\label{fig:convergence}}
\end{figure*}

\section{Numerical Experiment} \label{sec:experiment}
We use the {\tt athena++} code framework (\citealt{athena++}, submitted) to run a large suite of three dimensional hydrodynamic simulations on a static Cartesian mesh using an $\mathcal{E}= P/(\gamma - 1)$ equation of state with $\gamma = 5/3$. We adopt a standard, non-gravitating KHI setup that has cold dense gas moving relative to hot dilute gas with a shear velocity of $v_x = \vrel$. The two phases are in pressure equilibrium and initially smoothly connected following the procedure laid out by \cite{Lecoanet+16}. The velocity gradient is in the $\hat{z}$ direction. We seed the initial KHI with grid scale white noise and a sinusoidal $v_z$ perturbation with wavelength equal to the box size $L$ and an amplitude of $\vrel/25$ that declines exponentially with distance from the interface. The simulation domain is periodic in the $\hat{x}$ and $\hat{y}$ directions. In the $\hat{z}$ direction we enforce a boundary condition that holds the density $\rho$, pressure $P$, and streamwise velocity $v_x$ constant, and imposes a zero-gradient condition for $v_y$ and $v_z$. To ensure that evolution of the mixing layer is unaffected by the choice of vertical boundary condition we adopted a box that extends $10 L$ in the $\hat{z}$ direction, and $L$ in the $\hat{x}$ and $\hat{y}$ directions. We use a statically refined grid chosen to focus the resolution to the desired level within $-1.5 \leq z/L \leq 1.5$. The majority of our simulations are run with $\Delta x = L / 128$ in the most refined region, and we explore resolutions up to 4 times higher and 8 times lower.

We are interested in the case where cooling is dominated by the intermediate temperature gas, so we adopt a log-normal cooling curve $\Lambda(T)$ that by design peaks at the expected mixed phase temperature $\Tmix = \sqrt{\Thot \Tcold}$ \citep{BegelmanFabian90}. Although this choice sacrifices a degree of physical realism it simplifies the analysis, enhances our control over the experiments, and untethers our findings from specific physical regimes that would be imposed by choosing a particular cooling curve. This facilitates the application of our results to a range of environments. The functional form is specified by (i) the maximum value $\Lambda(\Tmix)$, which is adjusted to yield the desired cooling time at $\Tmix$, and (ii) the width, which is chosen so that the cooling curve at $\Tcold$ and $\Thot$ is ${\sim} 100$ less than at the peak. This closely approximates the cooling curve appropriate for the CGM, but is applicable to systems in the ISM, ICM, and protostellar jets because of their similar functional forms and the insensitivity of our results to the cooling curve width. For the remainder of the \emph{Letter} we use $\tc$ to refer to the cooling time of gas at $\Tmix$. Because the cooling rate scales as $\rho^2 \Lambda(T)$ the minimum cooling time is somewhat shorter than $\tc(\Tmix)$ and occurs at a temperature less that $\Tmix$. This introduces an order unity offset when comparing the simulations to \autoref{eq:final}.

Our parameter survey spans a broad range of the characteristic dimensionless numbers with $\chi = \rhoc/\rhoh$ ranging from 10 to 1000, $\mach = \vrel/\cshot$ from $10^{-1}$ to $10^{0.5}$, and $\xi = \tstc$ from $10^{-3}$ to $10^2$, as well as adiabatic/no cooling simulations with $\xi = 0$. In all cases we ran the simulations for at least 60 $\tsh$. Our fiducial simulation has $\xi = 10$, $\chi = 100$, $\mach=10^{-1/2}$, and $\Delta x = L/512$.

\section{Results} \label{sec:results}
 
\hyperref[fig:map]{Figure 1} visually demonstrates the properties of our numerical experiments of strongly cooling mixing layers, showing 2D slices of the 3D temperature, density, cooling time, pressure deviation, $v_x$, $v_y$, $v_z$, and turbulent Mach number $\macht=\vturb/\cs$ of our fiducial simulation. At this time, $t=30 \tsh$, the initial KHI has given way to fully developed turbulence---traced clearly by $v_y$---which promotes mixing and has broadened the shear velocity $v_x$ gradient. The turbulent mixing, however, is unable to broaden the temperature and density gradients because of the strong cooling that occurs as the phases mix. The cooling takes place entirely in a thin corrugated sheet that separates the hot and cold phase and leads to a net inflow from the hot phase. 

Although the cooling is rapid it is predominantly isobaric, as evidenced by the lack of a pressure decrement where the cooling is fastest. The pressure deviations correlate with the velocity fluctuations such that $\Delta P / \overline{P} \propto \macht^2$. This points to an essential concept that the rate of cooling, and therefore mass and momentum transfer, is limited by the turbulent mixing because the cooling does not increase the turbulent mixing when the cooling layer is well-resolved\footnote{In paper II we will present a model for the weak $\xi$ dependence of the turbulent velocities, highlighting in what (extreme) limits this breaks down, which is closely related to recent findings on whether thermally unstable clouds shatter \citep{Gronke+20b}.}.

In the presence of cooling there is a dichotomy between the thermal and momentum mixing layers. This arises because the contraction due to cooling offsets the broadening due to turbulent mixing of the temperature and density, but has (to first order) no effect on the shear velocity. \hyperref[fig:profiles]{Figure 2} shows the mass-weighted horizontally averaged temperature $\widetilde{T}$ (top) and shear velocity $\widetilde{v_x}$ (bottom) profiles at $t/\tsh = 30$. The profiles are shifted so the velocities equal zero at the same height. The z-location of the $\widetilde{v_x}=0$ point increases less in more rapidly cooling simulations (shown in the small colored ticks) because of the inflow ram pressure. The shape of the velocity profile is nearly independent of $\xi$ with minor deviations becoming clear in the higher $\chi$ simulations. The shape of the temperature profile, however, depends sensitively on the degree of cooling---becoming steeper in more rapidly cooling (higher $\xi$) simulations. 

Although the steepening of the average temperature profile is a hallmark of rapid cooling, the essential properties of the complex cooling surface are lost when horizontally averaged. The basis of the model presented in \autoref{sec:model} is that high enthalpy hot gas that flows into the mixing layer loses its thermal energy in a thin sheet with fractal properties. The lower left panel of \autoref{fig:Fractal_Dimension} shows the temperature isosurface defined by the locus where the cooling time is at its minimum. The surface is inherently rough and shows structure on all scales. 

We measure the fractal dimension by calculating how the isosurface area decreases when the temperature field is blurred (i.e. downsampled) on scale $\lambda$. Examples of the isosurface when blurred by $\lambda = 8,\, 16$ and $32\, \Delta x$, which corresponds to $\lambda = L/64,\, L/32,$ and $L/16$, are shown in the lower right panels. The top left panel shows quantitatively how the blurred isosurface area $A_\lambda$ scales with $\lambda$. The logarithmic derivative of this relationship directly corresponds to the fractal dimension and matches the $D=5/2$ prediction that $A_\lambda \propto \lambda^{-1/2}$ (\autoref{eq:area}). 

Finally, the top right panel of \autoref{fig:Fractal_Dimension} shows the thermal energy flux through a range of temperature isosurfaces. The flux is constant through the high temperature isosurfaces and drops precipitously once $T\lesssim \Tmix$ where the cooling rate increases dramatically. This validates the fundamental assumptions of our model that (i) enthalpy is conserved as hot gas is carried into the turbulent mixing layer until it has been mixed with enough cold gas to reach $\sim \Tmix$, at which point cooling rapidly drains the available thermal energy, which (ii) occurs in a thin corrugated sheet characterized by a fractal dimension of $D=5/2$.

The top panel of \autoref{fig:Edot} shows, for a single exemplary simulation, the nearly matching evolution of the normalized directly-measured inflow velocity $\vin$, total cooling rate $\edotcool$, cold phase mass growth rate $\mdot$, and cold phase acceleration $\pdot$. For each quantity, the normalization is simply based on the appropriate flux carried by the hot phase. The agreement of $\vin$ and $\edotcool$ demonstrates that, as predicted in \autoref{sec:model}, the enthalpy advection balances radiative losses, and that mass and momentum are carried into the cold phase along with the enthalpy. The flux predicted by the fractal cooling layer model (\autoref{eq:general}) given the measured turbulent velocity is also shown and accurately tracks the measured fluxes.

The middle panel of \autoref{fig:Edot} shows the cooling rate at all times for nearly 100 simulations versus the predicted scaling using the measured $\vturb$ in \autoref{eq:general} with $C (\Lturb/L)^{1/4} = 0.15$. The comparison with \autoref{eq:general}, which allows for weak evolution of $\vturb$ in time for any given simulation, demonstrates that the model captures the evolution of individual systems as well as the differences between systems.

The bottom panel of \autoref{fig:Edot} shows the measured average cooling rate from 20 to 40 $\tsh$ normalized by the predicted enthalpy flux (\autoref{eq:edot} and \autoref{eq:final}) for all simulations---spanning 4 orders of magnitude in $\xi$, and a broad range of $\chi$ and $\mach$. We adopt a coefficient $0.04$ that includes $\ft$ and the order unity constants in the expressions for $w$ and $A_w$ in \autoref{eq:w} and \autoref{eq:area}. The prediction correctly captures the dependence of $\edotcool$ on $\xi,\, \chi$, and $\mach$ in the rapid cooling limit ($\xi > 1$). The slowly cooling systems have not had enough time ($\gtrsim {\rm few}\,\tc$) to equilibrate, but it is likely that in more realistic environments they would first be disrupted \citep{Gronke+18}. \emph{The close agreement of our prediction and the experimental outcome demonstrates that the essential behavior of these complex and ubiquitous systems can be encapsulated by a power law relation of the three dimensionless numbers that describe the bulk properties.}

Finally, the top panels of \autoref{fig:convergence} show the resolution dependence of the median pressure-entropy phase diagrams of two rapidly cooling systems ($\xi = 10, 10^{3/2}$). Low resolution simulations exhibit substantial pressure dips at intermediate entropy where the cooling is most rapid, but as the resolution is increased the pressure dips vanish. Pressure dips are a result of numerical diffusion artificially broadening the cooling layer. The pressure dips in under-resolved simulations increase with $\xi$ and $\chi$, and can lead to spurious turbulent driving that is not present with higher resolutions \citep[possibly at play in][which had higher $\chi$ and $\xi$ and relatively low resolution]{Gronke+20}. Even though the phase structure depends strongly on the resolution, the total cooling, shown in the bottom panels of \autoref{fig:convergence}, is accurate to better than a factor of two for the lowest resolutions and is well converged for $\Delta x \lesssim L/128$.

\section{Discussion}\label{sec:discussion}

Many recent works have studied closely related problems, such as the turbulent mixing of slabs, sheets, and cylinders both without cooling \citep[\eg][]{Mandelker+19a}, and with cooling \citep[\eg][]{Ji+19, Mandelker+19b}, and the impact of cooling on ``cloud-crushing'' \citep{Scannapieco+15, Armillotta+16, Gronke+18, Gronke+20, Sparre+19, Li+20}. We now discuss some of these recent works in the context of our theory.

\cite{Ji+19} adopted a similar numerical setup and considered the balance of cooling with the advection of enthalpy from the hot phase, which also forms the basis of our model. Their analyses, however, focused on horizontally averaged quantities, which wipes out the essential fractal properties of the cooling layer. Because the surface is corrugated, but not entirely volume filling (i.e.,\, $D < 3$), horizontal averages combine the cooling and inert material. They treat the cooling volume as a flat sheet with area $L^2$ and a thickness set by the balance of diffusion and cooling, which misses the large increase in cooling volume from the fractal nature of the surface area (see Eqs.~\ref{eq:vin} and \ref{eq:area}). This led them to propose a different scaling of $\vin$ with $\tc$ from our result. \cite{Ji+19} attributed pressure dips to rapid cooling, but we instead suggest that pressure dips can instead be a signature of inadequate resolution.

\cite{Gronke+18,Gronke+20} demonstrated using radiative cloud crushing simulations that clouds that are large enough (such that the cloud crushing time $\chi^{1/2} \tsh$ is longer than the cooling time) grow in mass due to cooling at a rate corresponding to $\vin\propto \tc^{-1/4}$. This has since also been found in a shear flow set-up similar to ours \citep{Mandelker+19b}. These works, however, ascribe the inflow of high enthalpy hot gas into the mixing layer to the development of pressure gradients due to strong cooling \citep[as in][]{Ji+19}. Although the systems studied in these works are not exactly analogous to ours (clouds and cylinders as opposed to slabs) the underlying physics is likely the same, and we have demonstrated that the cooling is isobaric in fully resolved simulations. Rather than ascribing the driving of inflow to pressure gradients resulting from cooling, we instead believe that the inflow is fundamentally driven by turbulence. Shear creates the turbulence that mixes the layers at the interface, and this would be true regardless of cooling. We discuss this in more detail in Paper II.
Although these authors do not explicitly identify the additional $\vrel^{3/4}$ and $\chi^{3/8}$ dependence of $\vin$ (see \autoref{eq:general} and \autoref{fig:Edot}), there are some hints of this in their results. 

A limitation of our numerical experiment is its micro-scale scope. Meso-scale effects such as the expansion or destruction of the cold phase cannot be captured in our setup, and would require, e.g., cloud crushing or filament mixing simulations. The macro-scale environment may also impact how radiative mixing layers manifest in reality by introducing other length or time scales. For example, the background hot phase may be turbulent whereas we have assumed it to be laminar.

Our simulations and model do not include magnetic fields, viscosity, or conduction, which have been shown to change or suppress mixing and alter the phase structure when strong enough \citep[\eg][]{Armillotta+17,Berlok+19b}. We plan to investigate these effects in a future work, but are encouraged that \cite{Gronke+20} found the cold phase growth rate to be nearly independent of magnetic field strength and that \citet{Armillotta+16} found that condensation can occur in the presence of appreciable conduction.

In summary, our model for the fractal nature of the cooling surface in radiative turbulent mixing layers provides physical insight and a simple mathematical expression for the rate of energy loss to cooling as well as the mass and momentum transfer from the hot phase to the cold phase. Our model predicts that cold phase growth and entrainment driven by KHI is enhanced in environments with (i) high relative velocities, (ii) large density contrasts, and (iii) rapid cooling. This model accurately captures the behavior of our shear flow numerical experiments. It is expected to apply generally in scenarios where turbulent mixing promotes strong cooling, which is common in a broad range of astrophysical contexts, such as star forming regions, ISM, galactic winds, CGM, and ICM. 

\acknowledgments 
We are grateful to Eliot Quataert, Keaton Burns, Daniel Lecoanet, Peng Oh, Max Gronke, Suoqing Ji, and Chang-Goo Kim for stimulating discussions. DBF and ASJ are supported by the Simons Foundation through the Flatiron Institute. DBF thanks the Aspen Center for Physics supported by NSF PHY-1607611, where part of this work was completed, for its hospitality. The work of ECO was supported in part by award 510940 from the Simons Foundation.
GLB acknowledges support from NSF grants AST-1615955 and OAC-1835509 and NASA grant NNX15AB20G. 

\bibliographystyle{apj}
\bibliography{references}

\end{document}